\documentclass[10pt,conference]{IEEEtran}

\newtheorem{theorem}{Theorem}
\newtheorem{proposition}{Proposition}
\newtheorem{corollary}{Corollary}
\newtheorem{definition}{Definition}
\newtheorem{lemma}{Lemma}

\newcommand{\df}{\stackrel{\mbox{\scriptsize def}}{=}}

\newcommand{\rk}{\mathrm{rk}}
\newcommand{\wh}{\mathrm{w}_{\mbox{\tiny{H}}}}
\newcommand{\dr}{d_{\mbox{\tiny{R}}}}

\newcommand{\vspan}[1]{\left< #1 \right>}

\usepackage{amsmath, amssymb,cite}
\usepackage[dvips]{graphicx}

\begin{document}
\title{MacWilliams Identity for the Rank Metric}
\author{
\authorblockN{Maximilien Gadouleau and Zhiyuan Yan}
\authorblockA{Department of Electrical and Computer Engineering \\
Lehigh University, PA 18015, USA\\ E-mails: \{magc,
yan\}@lehigh.edu}}
\maketitle

\begin{abstract}
This paper investigates the relationship between the rank weight
distribution of a linear code and that of its dual code. The main
result of this paper is that, similar to the MacWilliams identity
for the Hamming metric, the rank weight distribution of any linear
code can be expressed as an analytical expression of that of its
dual code. Remarkably, our new identity has a similar form to the
MacWilliams identity for the Hamming metric. Our new identity
provides a significant analytical tool to the rank weight
distribution analysis of linear codes. We use a linear space based
approach in the proof for our new identity, and adapt this approach
to provide an alternative proof of the MacWilliams identity for the
Hamming metric. Finally, we determine the relationship between
moments of the rank distribution of a linear code and those of its
dual code, and provide an alternative derivation of the rank weight
distribution of maximum rank distance codes.
\end{abstract}

\section{Introduction}\label{sec:introduction}
The rank metric has attracted some attention due to its relevance to
wireless communications \cite{tarokh_98,lusina_it03}, public-key
cryptosystems \cite{gabidulin_lncs91}, and storage equipments (see,
for example, \cite{roth_it91}). Due to these applications, there is
a steady stream of work that focus on general properties of codes
with the rank metric
\cite{gabidulin_pit0185,roth_it91,babu_95,vasantha_gs99,manoj_report1002,vasantha_itw1002,vasantha_icads1202,sripati_isit03,kshevetskiy_isit05,gabidulin_isit05,loidreau_07}.
Despite these works, many open problems remain for rank metric
codes. For example, it is unknown how to derive the rank weight
distribution for any given linear code except when the code is a
maximum rank distance (MRD) code \cite{gabidulin_pit0185}. Besides
the minimum rank distance, the rank weight distribution is an
important property of any rank metric code, and determines its error
performance in applications.

In this paper, we investigate the rank weight properties of linear
codes. The main result of this paper is that, similar to the
MacWilliams identity for the Hamming metric, the rank weight
distribution of any linear code can be expressed as an analytical
expression of that of its dual code. Our new identity is a
significant analytical tool for both rank weight distribution and
hence error performance analysis of linear codes. To our best
knowledge, no similar result exists in the literature. It is also
remarkable that our new MacWilliams identity for the rank metric has
a similar form to that for the Hamming metric. Despite the
similarity, our new identity is proved using a different approach
based on linear spaces. Using the same approach, we give an
alternative proof of the MacWilliams identity for the Hamming
metric. Based on our new identity, we also derive an expression that
relates moments of the rank distribution of a linear code to those
of its dual code, and provide an alternative derivation for the rank
weight distribution of MRD codes.

The rest of the paper is organized as follows.
Section~\ref{sec:preliminaries} reviews necessary backgrounds in an
effort to make this paper self-contained.
Section~\ref{sec:star_prod_q-derivative} introduces the concepts of
$q$-product and $q$-derivative for homogeneous polynomials, and
investigates their properties. Using these tools,
Sections~\ref{sec:dual_v_rank} and~\ref{sec:theorem_rank} prove the
MacWilliams identity for the rank metric, and
Section~\ref{sec:moments_distrib} derives the relationship between
the moments of the rank distribution of a linear code and those of
its dual code. We also provide an alternative derivation of the rank
distribution of MRD codes in Section~\ref{sec:distribution_MRD}.
Some examples are provided in Section~\ref{sec:examples} to
illustrate our results. Finally,
Section~\ref{sec:macwilliams_hamming} adapts the approach in
Sections~\ref{sec:dual_v_rank} and~\ref{sec:theorem_rank} to provide
an alternative proof of the MacWilliams identity for the Hamming
metric. All the proofs have been omitted due to limited space, and
they will be presented at the conference.

\section{Preliminaries}\label{sec:preliminaries}
\subsection{Rank metric}\label{sec:rank_metric}
Consider an $n$-dimensional vector ${\bf x} = (x_0, x_1,\ldots,
x_{n-1}) \in \mathrm{GF}(q^m)^n$. Assume $\{\alpha_0, \alpha_1,
\ldots, \alpha_{m-1}\}$ is a basis set of GF$(q^m)$ over GF$(q)$,
then for $j=0, 1, \ldots, n-1$, $x_j$ can be written as $x_j =
\sum_{i=0}^{m-1} x_{i, j}\alpha_i$, where $x_{i, j} \in
\mbox{GF}(q)$ for $i=0, 1, \ldots, m-1$. Hence, $x_j$ can be
expanded to an $m$-dimensional column vector $(x_{0, j}, x_{1,
j},\ldots, x_{m-1, j})^T$ with respect to the basis set $\{\alpha_0,
\alpha_1, \ldots, \alpha_{m-1}\}$. Let ${\bf X}$ be the $m\times n$
matrix obtained by expanding all the coordinates of ${\bf x}$. That
is,
\begin{displaymath}
    {\bf X} = \left(
    \begin{array}{cccc}
        x_{0, 0} & x_{0, 1} & \ldots & x_{0, n-1}\\
        x_{1, 0} & x_{1, 1} & \ldots & x_{1, n-1}\\
        \vdots & \vdots & \ddots & \vdots\\
        x_{m-1, 0} & x_{m-1, 1} & \ldots & x_{m-1, n-1}
    \end{array}
    \right),
\end{displaymath}
where $x_j = \sum_{i=0}^{m-1} x_{i, j}\alpha_i$.  The \emph{rank
norm} of the vector ${\bf x}$ (over GF$(q)$), denoted as $\rk({\bf
x}|\mbox{GF}(q))$, is defined to be the rank of the matrix ${\bf X}$
over GF$(q)$, i.e., $\rk({\bf x}|\mbox{GF}(q)) \df
\mathrm{rank}({\bf X})$ \cite{gabidulin_pit0185}. In this paper, all
the ranks are over the base field GF$(q)$ unless otherwise
specified. To simplify notations, we denote the rank norm of ${\bf
x}$ as $\rk({\bf x})$ henceforth.

The rank norm of ${\bf x}$ is also the number of coordinates in
${\bf x}$ that are linearly independent over $\mathrm{GF}(q)$
\cite{gabidulin_pit0185}. The field $\mathrm{GF}(q^m)$ may be viewed
as an $m$-dimensional vector space over $\mathrm{GF}(q)$. The
coordinates of ${\bf x}$ thus span a linear subspace of
$\mathrm{GF}(q^m)$, denoted as $\mathfrak{S}({\bf x})$, and the rank
of ${\bf x}$ is the dimension of $\mathfrak{S}({\bf x})$.

For all ${\bf x}, {\bf y}\in \mathrm{GF}(q^m)^n$, it is easily
verified that $\dr({\bf x},{\bf y})\df \rk({\bf x} - {\bf y})$ is a
metric over GF$(q^m)^n$, referred to as the \emph{rank metric}
henceforth \cite{gabidulin_pit0185}. The {\em minimum rank distance}
of a code, denoted as $d_{\mbox{\tiny R}}$, is simply the minimum
rank distance over all possible pairs of distinct codewords.

\subsection{The Singleton bound and MRD codes}\label{sec:singleton}
The minimum rank distance of a code can be specifically bounded.
First, the minimum rank distance $d_{\mbox{\tiny R}}$ of a code of
length $n$ over $\mathrm{GF}(q^m)$ is obviously bounded above by
$\min\{m,n\}$. Codes that satisfy $d_{\mbox{\tiny R}} = m$ are
studied in \cite{manoj_report1002}. Also, it can be shown that
$d_{\mbox{\tiny R}} \leq d_{\mbox{\tiny H}}$
\cite{gabidulin_pit0185}, where $d_{\mbox{\tiny H}}$ is the minimum
Hamming distance of the same code. Due to the Singleton bound on the
minimum Hamming distance of block codes \cite{macwilliams_77}, the
minimum rank distance of a block code of length $n$ ($n\leq m$) and
cardinality $M$ over $\mathrm{GF}(q^m)$ thus satisfies
\begin{equation}\label{eq:singleton1}
    d_{\mbox{\tiny R}} \leq n-\log_{q^m}M+1.
\end{equation}
As in \cite{gabidulin_pit0185}, we refer to codes that achieve the
equality in Eq.~(\ref{eq:singleton1}) as MRD codes. It is also shown
that the dual of any MRD code is also an MRD code
\cite{gabidulin_pit0185}. Clearly MRD codes are the counterparts of
maximum distance separable (MDS) codes.

\subsection{Weight enumerator and Hadamard
transform}\label{sec:hadamard}

We restrict our attention to the Hamming metric and the rank metric
only henceforth in this paper.

\begin{definition}[Weight function]\label{def:f_w}
Let $d$ be a metric over $\mathrm{GF}(q^m)^n$, and define
$\mathrm{w}({\bf v}) = d({\bf 0},{\bf v})$ as a weight over
$\mathrm{GF}(q^m)^n$. Suppose ${\bf v} \in \mathrm{GF}(q^m)^n$ has
weight $r$, then the weight function of ${\bf v}$ is defined as
$f_{\mathrm{w}}({\bf v}) = y^{r}x^{n-r}$.
\end{definition}

We shall henceforth denote the Hamming weight function and the rank
weight function as $f_{\mbox{\tiny{H}}}$ and $f_{\mbox{\tiny{R}}}$
respectively. Note that $n$ is the maximum weight for both
$f_{\mbox{\tiny{H}}}$ and $f_{\mbox{\tiny{R}}}$.

\begin{definition}\label{def:W_C}
Let $\mathcal{C}$ be a code of length $n$ over $\mathrm{GF}(q^m)$.
Suppose there are $A_i$ codewords in $\mathcal{C}$ with weight $i$,
then the weight enumerator of $\mathcal{C}$, denoted as
$W_\mathcal{C}(x,y)$, is defined as
$$
    W^{\mathrm{w}}_\mathcal{C}(x,y) \df \sum_{{\bf v} \in \mathcal{C}}f_{\mathrm{w}}({\bf v}) = \sum_{i=0}^n A_i y^{i}x^{n-i}.
$$
\end{definition}

\begin{definition}[Hadamard transform
\cite{macwilliams_77}]\label{def:hadamard} Let $\mathbb{C}$ be the
field of complex numbers. Let $a \in \mathrm{GF}(q^m)$ and let
$\{1,\alpha_1,\ldots,\alpha_{m-1} \}$ be a basis set of
$\mathrm{GF}(q^m)$. We thus have $a = a_0 + a_1 \alpha_1 + \ldots +
a_{m-1}\alpha_{m-1}$. Finally, let $\zeta \in \mathbb{C}$ be a
primitive $q$-th root of unity. We define $\chi(a) \df \zeta^{a_0}$.
For a mapping $f$ from $F$ to $\mathbb{C}$, the {\em Hadamard
transform} of $f$, denoted as $\hat{f}$, is given by
\begin{equation}\label{eq:hadamard}
    \hat{f}({\bf v}) \df \sum_{{\bf u} \in F} \chi({\bf u} \cdot {\bf
    v}) f({\bf u}).
\end{equation}
\end{definition}

\subsection{Notations}\label{sec:notations}
In order to simplify notations, we shall occasionally denote the
vector space $\mathrm{GF}(q^m)^n$ as $F$. We denote the number of
vectors of rank $u$ ($0 \leq u \leq \min\{m,n\}$) in
$\mathrm{GF}(q^m)^n$ as $N_u(q^m,n)$. It can be shown that
$N_u(q^m,n) = {n \brack u} \alpha(m,u)$, where $\alpha(m,u)$ is
defined as follows: $\alpha(m,0) = 1$ and $\alpha(m,u) =
\prod_{i=0}^{u-1}(q^m-q^i)$ for $u \geq 1$. The ${n \brack u}$ term
is the Gaussian binomial~\cite{andrews}, defined as ${n \brack u} =
\alpha(n,u)/\alpha(u,u)$. Note that ${n \brack u}$ is the number of
$u$-dimensional linear subspaces of $\mathrm{GF}(q)^n$. We also
define $\beta(m,0) \df 1$ and $\beta(m,u) \df \prod_{i=0}^{u-1} {m-i
\brack 1}$ for $u>0$, which are used in
Section~\ref{sec:star_prod_q-derivative}. These terms are closely
related to the Gaussian binomial: $\beta(m,u) = {m \brack u}
\beta(u,u)$ and $\beta(m+u,m+u) = {m+u \brack u} \beta(m,m)
\beta(u,u)$.

\section{MacWilliams identity for the rank
metric}\label{sec:macwilliams_rank}

\subsection{$q$-product and $q$-derivative of homogeneous polynomials}
\label{sec:star_prod_q-derivative}

\begin{definition}[$q$-product]\label{def:star_prod}
Let $a(x,y;m) = \sum_{i=0}^r a_{i}(m) y^i x^{r-i}$ and $b(x,y;m) =
\sum_{j=0}^s b_{j}(m) y^j x^{s-j}$ be two homogeneous polynomials in
$x$ and $y$ of degrees $r$ and $s$ respectively with coefficients
$a_i(m)$ and $b_j(m)$ respectively. $a_i(m)$ and $b_j(m)$ for $i,j
\geq 0$ in turn are real functions of $m$, and are assumed to be
zero unless otherwise specified. The $q$-product of $a(x,y;m)$ and
$b(x,y;m)$ is defined to be the homogeneous polynomial of degree
$(r+s)$ $c(x,y;m) \df a(x,y;m)
* b(x,y;m) = \sum_{u=0}^{r+s} c_{u}(m) y^u x^{r+s-u}$, with
$$
    c_{u}(m) = \sum_{i=0}^u q^{is} a_{i}(m) b_{u-i}(m-i).
$$

For $n \geq 0$ the $n$-th $q$-power of $a(x,y;m)$ is defined
recursively: $a(x,y;m)^{[0]} = 1$ and $a(x,y;m)^{[n]} =
a(x,y;m)^{[n-1]}*a(x,y;m)$ for $n \geq 1$.
\end{definition}

To illustrate the $q$-product, we provide some examples of the
$q$-product. We have $x * y = yx$, $y * x = qyx$, $yx * x = q yx^2$,
and $yx * (q^m-1)y = (q^m-q)y^2x$. Note that $x * y \ne y * x$. It
is easy to verify that the $q$-product is in general
non-commutative. However, it is commutative for some special cases.
\begin{lemma}\label{lemma:properties_star_prod}
Suppose $a(x,y;m) = a$ is a constant independent from $m$, then
$a(x,y;m)*b(x,y;m) = b(x,y;m)*a(x,y;m) = ab(x,y;m)$. Also, if
$\deg[c(x,y;m)] = \deg[a(x,y;m)]$, then
$[a(x,y;m)+c(x,y;m)]*b(x,y;m) = a(x,y;m)*b(x,y;m) +
c(x,y;m)*b(x,y;m)$, and $b(x,y;m) * [a(x,y;m)+c(x,y;m)] =
b(x,y;m)*a(x,y;m) + b(x,y;m)*c(x,y;m)$.
\end{lemma}

The homogeneous polynomials $a_l(x,y;m) \df [x+(q^m-1)y]^{[l]}$ and
$b_l(x,y;m) \df (x-y)^{[l]}$ turn out to be very important to our
derivations below. The following lemma provides the analytical
expressions of $a_l(x,y;m)$ and $b_l(x,y;m)$.
\begin{lemma}\label{lemma:special_prod}
For $i \geq 0$, $\sigma_i \df \frac{i(i-1)}{2}$. For $l\geq 0$, we
have $y^{[l]} = q^{\sigma_l}y^l$ and $x^{[l]} = x^l$. Furthermore,
\begin{eqnarray}
    \label{eq:x+y^s}
    a_l(x,y;m) = \sum_{u=0}^l {l \brack u} \alpha(m,u) y^u
    x^{l-u},\\
    \label{eq:x-y^r}
    b_l(x,y;m) = \sum_{u=0}^l {l \brack u} (-1)^u q^{\sigma_u} y^u x^{l-u}.
\end{eqnarray}
\end{lemma}
Note that $a_l(x,y;m)$ is the rank weight enumerator of
$\mathrm{GF}(q^m)^l$.

\begin{definition}[$q$-transform]\label{def:star_transform}
We define the $q$-transform of $a(x,y;m)= \sum_{i=0}^r a_{i}(m) y^i
x^{r-i}$ as the homogeneous polynomial $\bar{a}(x,y;m)= \sum_{i=0}^r
a_{i}(m) y^{[i]}* x^{[r-i]}$.
\end{definition}

\begin{definition}[$q$-derivative
\cite{gasper_book04}]\label{def:q-derivative} For $q \geq 2$, the
$q$-derivative for $x \neq 0$ of a real-valued function $f(x)$ is
defined as
$$
    f^{(1)}(x) \df \frac{f(qx)-f(x)}{(q-1)x}.
$$
The $q$-derivative operator is linear. For $\nu \geq 0$, we shall
denote the  partial $\nu$-th $q$-derivative of $f(x,y)$ (with
respect to $x$) as $f^{(\nu)}(x,y)$. The $0$-th $q$-derivative of
$f(x,y)$ is defined to be $f(x,y)$ itself.
\end{definition}

\begin{lemma}\label{lemma:q-derivative_x}
For $\nu \leq n$, the $\nu$-th $q$-derivative of the function $x^n$
is given by $\beta(n,\nu)x^{n-\nu}$. Also, the $\nu$-th
$q$-derivative of $f(x,y)= \sum_{i=0}^r f_i y^i x^{r-i}$ is given by
$f^{(\nu)}(x,y) = \sum_{i=\nu}^r f_i \beta(i,\nu) x^{i-\nu}$.
\end{lemma}

\begin{lemma}[Leibniz rule]\label{lemma:Leibniz_x}
For two homogeneous polynomials $f(x,y) = \sum_{i=0}^r f_i y^i
x^{r-i}$ and $g(x,y) = \sum_{j=0}^s g_j y^j x^{s-j}$ with degrees
$r$ and $s$ respectively, the $\nu$-th ($\nu \geq 1$) $q$-derivative
of their $q$-product is given by
\begin{eqnarray}
    (f(x,y)*g(x,y))^{(\nu)} &=& \sum_{l=0}^{\nu} {\nu \brack l}
    q^{(\nu-l)(r-l)} \cdots \nonumber \\
    \label{eq:leibniz_nu}
    &\cdots& f^{(l)}(x,y)*g^{(\nu-l)}(x,y).
\end{eqnarray}
\end{lemma}

Next, we derive the $q$-derivatives of $a_l(x,y;m) =
[x+(q^m-1)y]^{[l]}$ and $b_l(x,y;m) = (x-y)^{[l]}$.

\begin{lemma}\label{lemma:special_q-d}
For $\nu \leq l$ we have
\begin{eqnarray}
    a_l^{(\nu)}(x,y;m) &=& \beta(l,\nu) a_{l-\nu}(x,y;m)\\
    b_l^{(\nu)}(x,y;m) &=& \beta(l,\nu) b_{l-\nu}(x,y;m).
\end{eqnarray}
\end{lemma}

\subsection{The dual of a vector}\label{sec:dual_v_rank}
As an important step toward our main result, we derive the rank
weight enumerator of $\vspan{{\bf v}}^\perp$, where ${\bf v} \in
\mathrm{GF}(q^m)^{n}$ is an arbitrary vector and $\vspan{{\bf v}}\df
\left\{a{\bf v}: a \in \mathrm{GF}(q^m)\right\}$. It is remarkable
that the rank weight enumerator of $\vspan{{\bf v}}^\perp$ depends
on only the rank of ${\bf v}$.

\begin{definition}\label{def:elementary_extension}
For $s \geq 1$ the $s$-th order ${\bf B}$-elementary extension of an
$(n,k)$ linear code $\mathcal{C}_0$ is the $(n+s,k+s)$ linear code
defined as $\mathcal{C}_s \df \{ (c_0,\ldots,c_{n+s-1}) \in
\mathrm{GF}(q^m)^{n+s} | (c_0,\ldots,c_{n-1}) -
(c_n,\ldots,c_{n+s-1}){\bf B} \in \mathcal{C}_0\}$, where ${\bf B}$
is an $s \times n$ matrix over $\mathrm{GF}(q)$. The $0$-th order
elementary extension of $\mathcal{C}_0$ is defined to be
$\mathcal{C}_0$ itself.
\end{definition}

\begin{lemma}\label{lemma:elementary_extension}
Let $\mathcal{C}_0$ be an $(n,k)$ linear code over
$\mathrm{GF}(q^m)$ with generator matrix ${\bf G}_0$ and
parity-check matrix ${\bf H}_0$. The $s$-th order ${\bf
B}$-elementary extension of $\mathcal{C}_0$ is the $(n+s,k+s)$
linear code $\mathcal{C}_s$ over $\mathrm{GF}(q^m)$ with a generator
matrix ${\bf G}_s  = \left( \begin{array}{c|c} {\bf G}_0 & {\bf 0}\\
\hline {\bf B} & {\bf I}_s \end{array}\right)$ and a parity-check
matrix ${\bf H}_s = \left(\begin{array}{c|c} {\bf H}_0  & -{\bf H}_0
{\bf B}^T \end{array}\right)$.
\end{lemma}

\begin{corollary}\label{cor:rk_dual_vector}
Suppose ${\bf v} = (v_0,\ldots,v_{n-1}) \in \mathrm{GF}(q^m)^n$ has
rank $r \geq 1$. Then $\mathcal{L} = \vspan{{\bf v}}^{\perp}$ is
equivalent to the $(n-r)$-th order elementary extension of an
$(r,r-1)$ linear code with $d_{\mbox{\tiny R}}=2$.
\end{corollary}
It is easy to verify that the $(r,r-1)$ code with $d_{\mbox{\tiny
R}}=2$ is actually an MRD code as defined in
Section~\ref{sec:singleton}.

We hence derive the rank weight enumerator of an $(r,r-1,2)$ MRD
code. Note that the rank weight distribution of MRD codes has been
derived in \cite{gabidulin_pit0185}. However, we will use our
results to give an alternative derivation of the rank weight
distribution of MRD codes later, and thus we shall not use the
result in \cite{gabidulin_pit0185} here.

\begin{lemma}\label{lemma:rec_Arr}
For $r \geq 1$, suppose ${\bf v}_r = (v_0,\ldots,v_{r-1}) \in
\mathrm{GF}(q^m)^r$ has rank $r \leq m$. Then the number of vectors
in $\mathcal{L}_r = \vspan{{\bf v}_r}^{\perp}$ with rank $r$,
denoted as $A_{r,r}$, depends on only $r$ and satisfies $A_{r,r} =
\alpha(m,r-1) - q^{r-1}A_{r-1,r-1}$. Furthermore, the rank weight
enumerator of $\mathcal{L}_r$ is given by
$$
    W_{\mathcal{L}_r}^{\mbox{\tiny R}}(x,y) = q^{-m}\left\{\left[x+(q^m-1)y\right]^{[r]} +
    (q^m-1)(x-y)^{[r]}\right\}.
$$\end{lemma}

The following lemma relates the rank weight enumerator of a code to
that of any of its $s$-th order elementary extensions.

\begin{lemma}\label{lemma:rk_Asu1}
Let $\mathcal{C}_0 \subseteq \mathrm{GF}(q^m)^r$ be a linear code
with rank weight enumerator $W_{\mathcal{C}_0}^{\mbox{\tiny
R}}(x,y)$, and for $s \geq 0$, let $W_{\mathcal{C}_s}^{\mbox{\tiny
R}}(x,y)$ be the rank weight enumerator of its $s$-th order ${\bf
B}$-elementary extension ${\mathcal{C}_s}$. Then
$W_{\mathcal{C}_s}^{\mbox{\tiny R}}(x,y)$ does not depend on ${\bf
B}$ and is given by
\begin{equation}\label{eq:rk_Asu}
    W_{\mathcal{C}_s}^{\mbox{\tiny R}}(x,y) = W_{\mathcal{C}_0}^{\mbox{\tiny R}}(x,y) *
    \left[x+(q^m-1)y\right]^{[s]}.
\end{equation}
\end{lemma}

Combining Corollary~\ref{cor:rk_dual_vector},
Lemma~\ref{lemma:rec_Arr}, and Lemma~\ref{lemma:rk_Asu1}, the rank
weight enumerator of $\vspan{{\bf v}}^\perp$ can be determined at
last.

\begin{proposition}\label{prop:rk_W_L}
For ${\bf v} \in \mathrm{GF}(q^m)^n$ with rank $r \geq 0$, the rank
weight enumerator of $\mathcal{L} = \vspan{{\bf v}}^{\perp}$ depends
on only $r$, and is given by
\begin{eqnarray}
    \nonumber
    W_\mathcal{L}^{\mbox{\tiny R}}(x,y) &=& q^{-m} \left\{ \left[x+(q^m-1)y\right]^{[n]} + (q^m-1)\right. \cdots\\
    & \cdots & \left. (x-y)^{[r]} * \left[x+(q^m-1)y\right]^{[n-r]} \right\}.
\end{eqnarray}
\end{proposition}

\subsection{MacWilliams identity} \label{sec:theorem_rank}

Using the results shown in Section~\ref{sec:dual_v_rank}, we now
derive the MacWilliams identity for rank metric codes.

\begin{lemma}\label{lemma:hadamard}
Suppose that for all $\lambda \in \mathrm{GF}(q^m)^*$ and all ${\bf
u} \in F$, we have $\mathrm{w}(\lambda {\bf u}) = \mathrm{w}({\bf
u})$. For ${\bf v} \in \mathrm{GF}(q^m)^n$, let us denote
$\vspan{{\bf v}}^{\perp}$ as $\mathcal{L}$. Then the Hadamard
transform of the weight function $f_{\mathrm{w}}$, denoted as
$\hat{f}_{\mathrm{w}}$, satisfies
\begin{equation}\label{eq:fhat_wl}
    W^{\mathrm{w}}_\mathcal{L}(x,y) = q^{-m} \left[W^{\mathrm{w}}_F(x,y) + (q^m-1)\hat{f}_{\mathrm{w}}({\bf v}) \right].
\end{equation}
\end{lemma}

\begin{lemma}\label{cor:f_R_hat}
Suppose ${\bf v} \in \mathrm{GF}(q^m)^n$ has rank $r$. Then the
Hadamard transform of the rank weight function is given by
\begin{equation}\label{eq:f_rk}
    \hat{f}_{\mbox{\tiny{R}}}({\bf v}) = (x-y)^{[r]}* \left[x+(q^m-1)y\right]^{[n-r]}.
\end{equation}
\end{lemma}

Let $\mathcal{C}$ be an $(n,k)$ linear code over $\mathrm{GF}(q^m)$,
and let $W_\mathcal{C}^{\mbox{\tiny R}}(x,y) = \sum_{i=0}^n A_i
y^ix^{n-i}$ be its rank weight enumerator and
$W_{\mathcal{C}^{\perp}}^{\mbox{\tiny R}}(x,y) = \sum_{j=0}^n B_j
y^j x^{n-j}$ be the rank weight enumerator of its dual code
$\mathcal{C}^{\perp}$.

\begin{theorem}\label{th:MacWilliams}
For any linear code $\mathcal{C}$ and its dual code
$\mathcal{C}^{\perp}$,
\begin{equation}\label{eq:macwilliams}
    W_{\mathcal{C}^{\perp}}^{\mbox{\tiny R}}(x,y) = \frac{1}{|\mathcal{C}|}
    {\bar W}_\mathcal{C}^{\mbox{\tiny R}}\left(x+(q^m-1)y,x-y\right),
\end{equation}
where ${\bar W}_\mathcal{C}^{\mbox{\tiny R}}$ is the $q$-transform
of $W_\mathcal{C}^{\mbox{\tiny R}}$. Equivalently,
\begin{equation}
    \sum_{j=0}^n B_j y^j x^{n-j} = q^{m(k-n)}\sum_{i=0}^n A_i
    (x-y)^{[i]}* \left[x+(q^m-1)y\right]^{[n-i]}.
\end{equation}
\end{theorem}

Also, $B_j$'s can be explicitly expressed in terms of $A_i$'s.

\begin{corollary}\label{cor:krawtchouk}
We have
\begin{equation}
    B_j = \frac{1}{|\mathcal{C}|} \sum_{i=0}^n A_i P_j(i;m,n),
\end{equation}
where
\begin{equation}
    P_j(i;m,n) \df \sum_{l=0}^j {i \brack l}{n-i \brack j-l}(-1)^l
    q^{\sigma_l}q^{l(n-i)}\alpha(m-l,j-l).
\end{equation}
\end{corollary}

\subsection{Moments of the rank distribution}\label{sec:moments_distrib}

Next, we investigate the relationship between moments of the rank
distribution of a linear code and those of its dual code. Our
results parallel those in \cite[p. 131]{macwilliams_77}.

First, applying Theorem~\ref{th:MacWilliams} to
$\mathcal{C}^{\perp}$, we obtain
\begin{equation}\label{eq:before_nu}
    \sum_{i=0}^n A_i y^i x^{n-i} = q^{m(k-n)} \sum_{j=0}^n B_j b_j(x,y;m)*a_{n-j}(x,y;m).
\end{equation}
By $q$-differentiating Eq.~(\ref{eq:before_nu}) $\nu$ times with
respect to $x$ and using the Leibniz rule in
Lemma~\ref{lemma:Leibniz_x} as well as the results in
Lemma~\ref{lemma:special_q-d}, we obtain
\begin{proposition}\label{prop:bm_x}
For $0 \leq \nu \leq n$,
\begin{equation}\label{eq:bm_x}
     \sum_{i=0}^{n-\nu} {n-i \brack \nu} A_i = q^{m(k-\nu)}
     \sum_{j=0}^{\nu} {n-j \brack n-\nu} B_j.
\end{equation}
\end{proposition}

As in \cite{macwilliams_77}, we refer to the left hand side of
Eq.~(\ref{eq:bm_x}) as moments of the rank distribution of
$\mathcal{C}$. We remark that the cases where $\nu=0$ and $\nu=n$
are trivial. Also, Proposition~\ref{prop:bm_x} can be simplified if
$\nu$ is less than the minimum distance of the dual code.

\begin{corollary}\label{cor:binomial_moment_x}
Let $\dr'$ be the minimum rank distance of $\mathcal{C}^{\perp}$. If
$\nu < \dr'$, then
\begin{equation}
    \sum_{i=0}^{n-\nu} {n-i \brack \nu} A_i = q^{m(k-\nu)}{n \brack
    \nu}.
\end{equation}
\end{corollary}

\subsection{Rank distribution of MRD codes}
\label{sec:distribution_MRD} The rank distribution of MRD codes was
first given in \cite{gabidulin_pit0185}. Based on our results in
Section~\ref{sec:moments_distrib}, we provide an alternative
derivation of the rank distribution of MRD codes. In this
subsection, we assume $n\leq m$.

First, we obtain the following results necessary for our alternative
derivation of the rank distribution.
\begin{lemma}\label{lemma:gaussian_transform}
Let $\{a_j\}_{j=0}^l$ and $\{b_i\}_{i=0}^l$ be two sequences of real
numbers. Suppose that for $0 \leq j \leq l$ we have $a_j =
\sum_{i=0}^j {l-i \brack l-j} b_i$. Then for $0 \leq i \leq l$ we
have $b_i = \sum_{j=0}^i (-1)^{i-j}q^{\sigma_{i-j}} {l-j \brack
l-i}a_j$.
\end{lemma}

Based on Corollary~\ref{cor:binomial_moment_x} and using
Lemma~\ref{lemma:gaussian_transform}, we can derive the rank
distribution of MRD codes when $n\leq m$:
\begin{proposition}[Rank distribution of MRD
codes]\label{prop:distrib_MRD} Let $\mathcal{C}$ be an $(n,k,\dr)$
MRD code over $\mathrm{GF}(q^m)$ $(n \leq m)$, and let
$W_\mathcal{C}^{\mbox{\tiny R}}(x,y) = \sum_{i=0}^n A_i y^i x^{n-i}$
be its rank weight enumerator. We then have $A_0 = 1$ and for $0
\leq i \leq n-\dr$,
\begin{equation}
    A_{\dr+i} = {n \brack \dr+i} \sum_{j=0}^i (-1)^{i-j}q^{\sigma_{i-j}} {\dr+i \brack
    \dr+j} \left(q^{m(j+1)}-1\right).
\end{equation}
\end{proposition}
We remark that the above rank distribution is consistent with that
derived by Gabidulin in \cite{gabidulin_pit0185}.

\subsection{Examples}\label{sec:examples}
In this section, we illustrate Theorem~\ref{th:MacWilliams} and
Proposition~\ref{prop:bm_x} using some examples. For $m\geq 2$,
consider the $(3,2)$ linear code $\mathcal{C}_1$ over
$\mathrm{GF}(q^m)$ with generator matrix
$$
    {\bf G}_1 = \left(\begin{array}{ccc}
                1&\alpha&1\\
                1&\alpha&0
                \end{array}\right),
$$
where $\alpha$ is a primitive element of $\mathrm{GF}(q^m)$. It can
be verified that the rank weight enumerator of $\mathcal{C}_1$ is
given by $W_{\mathcal{C}_1}^{\mbox{\tiny R}}(x,y) = x^{3} +
(q^m-1)yx^{2} + q^2(q^m-1)y^{2}x + (q^m-q^2)(q^m-1)y^{3}.$ Applying
Theorem~\ref{th:MacWilliams}, we obtain
$W_{\mathcal{C}_1^{\perp}}^{\mbox{\tiny R}}(x,y) = x^{3} +
(q^m-1)y^{2}x$. We can verify by hand that
$W_{\mathcal{C}_1^{\perp}}^{\mbox{\tiny R}}(x,y)$ is indeed the rank
weight enumerator of $\mathcal{C}_1^{\perp}$, which has a generator
matrix ${\bf H}_1 = (\begin{array}{ccc}-\alpha&1&0\end{array})$. For
$\mathcal{C}_1$, both sides of (\ref{eq:bm_x}) are given by
$q^{2m}$, $q^m{3 \brack 1}$, $(q^m-1+{3 \brack 1})$, and $1$ for
$\nu=0, 1, 2, 3$ respectively. Note that the results hold when $m=2<
n=3$.

For $m \geq 4$, let us now consider the $(4,2)$ code $\mathcal{C}_2$
over $\mathrm{GF}(q^m)$ with the following generator matrix
$$
    {\bf G}_2 = \left(\begin{array}{cccc}
                1&\alpha&\alpha ^2&\alpha ^3\\
                1&\alpha ^q&\alpha ^{2q}&\alpha ^{3q}
                \end{array}\right).
$$
$\mathcal{C}_2$ is actually a $(4, 2)$ MRD code with $\dr=3$. Hence,
its dual code $\mathcal{C}_2^{\perp}$ is also a $(4, 2)$ MRD code
with $\dr=3$. The rank weight enumerators of both $\mathcal{C}_2$
and $\mathcal{C}_2^{\perp}$ can be readily obtained using
Proposition~\ref{prop:distrib_MRD}, and they are given by
$W_{\mathcal{C}_2}^{\mbox{\tiny
R}}(x,y)=W_{\mathcal{C}_2^{\perp}}^{\mbox{\tiny R}}(x,y) = x^{4} +
{4 \brack 1} (q^m-1) y^{3}x^{1} + \left\{q^{2m}-1-{4 \brack 1}
(q^m-1)\right\} y^{4}$. It can be verified that
$W_{\mathcal{C}_2}^{\mbox{\tiny R}}(x,y)$ and
$W_{\mathcal{C}_2^{\perp}}^{\mbox{\tiny R}}(x,y)$ satisfy
Theorem~\ref{th:MacWilliams}.  For $\mathcal{C}_2$, it can also be
verified that both sides of (\ref{eq:bm_x}) are $q^{2m},{4 \brack
1}q^m, {4 \brack 2}, {4 \brack 1}$, and $1$ for $\nu=0, 1, \cdots,
4$ respectively.

Finally, consider the $(7,4)$ code $\mathcal{C}_3$ over
$\mathrm{GF}(2^4)$ with the following generator matrix
$$
    {\bf G}_3 = \left(\begin{array}{ccccccc}
    1 & 0 & 0 & 0 & \beta^3   & \beta^6 & \beta^{12}\\
    0 & 1 & 0 & 0 & \beta^6 & \beta^{12} & 0\\
    0 & 0 & 1 & 0 & \beta^{12} & 0        & \beta^3\\
    0 & 0 & 0 & 1 & 0        & \beta^3   & \beta^6
    \end{array}\right),
$$
where $\beta$ is a primitive element of $\mathrm{GF}(2^4)$. Its rank
weight enumerator is given by $ W_{\mathcal{C}_3}^{\mbox{\tiny
R}}(x,y) = x^{7} + 105 y^{2}x^{5} + 7350 y^{3}x^{4} + 58080
y^{4}x^{3}$, Theorem~\ref{th:MacWilliams} indicates that the rank
weight enumerator of its dual code is given by
$W_{\mathcal{C}_3^{\perp}}^{\mbox{\tiny R}}(x,y) = x^{7} + 465
y^{3}x^{4} + 3630 y^{4}x^{3}$, which can be verified using
exhaustive search. It can also be verified that both sides of
(\ref{eq:bm_x}) for $\mathcal{C}_3$ are $2^{16},520192, 682752,
196416, 22416, 2772, 127$, and $1$ for $\nu=0, 1, \cdots, 7$
respectively.

\section{MacWilliams identity for the Hamming
metric}\label{sec:macwilliams_hamming}

In this section, we adapt the approach used in our proof of
Theorem~\ref{th:MacWilliams} to provide an alternative proof of the
MacWilliams identity for the Hamming metric. We first derive the
Hamming weight enumerator of $\vspan{{\bf v}}^\perp$, where ${\bf
v}$ is an arbitrary vector. Then, using this result and properties
of the Hadamard transform, we obtain the MacWilliams identity for
the Hamming metric.

\begin{definition}\label{def:coordinate_extension}
For $s \geq 1$, the $s$-th order coordinate extension of an $(n,k)$
linear code $\mathcal{C}_0$ is defined as the $(n+s,k+s)$ code
$\mathcal{C}_s \df \{(c_0,\ldots,c_{n+s-1}) \in
\mathrm{GF}(q^m)^{n+s} | (c_0,\ldots,c_{n-1}) \in \mathcal{C}_0 \}$.
The $0$-th order coordinate extension of $\mathcal{C}_0$ is defined
as $\mathcal{C}_0$ itself.
\end{definition}
We remark that the $s$-th order coordinate extension is a special
case of the $s$-th order ${\bf B}$-elementary extension  with ${\bf
B}={\bf 0}$.
\begin{lemma}\label{lemma:coordinate_extension}
Let $\mathcal{C}_0$ be an $(n,k)$ linear code over
$\mathrm{GF}(q^m)$, with a generator matrix ${\bf G}_0$ and a
parity-check matrix ${\bf H}_0$. Then $\mathcal{C}_s$ over
$\mathrm{GF}(q^m)$ has a generator matrix ${\bf G}_s = \left(
\begin{array}{c|c} {\bf G}_0 & {\bf 0}\\ \hline {\bf 0} & {\bf I}_s \end{array}\right)$
and a parity-check matrix ${\bf H}_s = \left(\begin{array}{c|c} {\bf
H}_0  & {\bf 0} \end{array}\right)$.
\end{lemma}

\begin{corollary}\label{cor:wh_dual_vector}
Suppose ${\bf v} \in \mathrm{GF}(q^m)^n$ has Hamming weight $r \geq
1$. Then $\mathcal{L} = \vspan{{\bf v}}^{\perp}$ is equivalent to
the $(n-r)$-th order coordinate extension of an $(r,r-1,2)$ MDS
code.
\end{corollary}

We hence derive the Hamming weight distribution of an $(r,r-1,2)$
MDS code. Note that \cite{macwilliams_77} gives the Hamming weight
distribution of all MDS codes. However, that proof relies on the
MacWilliams identity, and thus may not be used here.

\begin{lemma}\label{lemma:wh_MDS}
Suppose ${\bf v}_r = (v_0,\ldots,v_{r-1}) \in \mathrm{GF}(q^m)^r$
has Hamming weight $r$. Then $\mathcal{L}_r = \vspan{{\bf
v}_r}^{\perp}$ is an $(r,r-1,2)$ MDS code whose weight enumerator
does not depend on ${\bf v}_r$ and is given by
$$
    W_{\mathcal{L}_r}^{\mbox{\tiny H}}(x,y) = q^{-m}\left\{\left[x+(q^m-1)y\right]^r + (q^m-1)(x-y)^r\right\}.
$$
\end{lemma}

The following lemma relates the Hamming weight enumerator of a code
to that of its $s$-th order coordinate extension.

\begin{lemma}\label{lemma:wh_Ws}
Let $\mathcal{C}_0 \subseteq \mathrm{GF}(q^m)^r$ be a linear code
with Hamming weight enumerator $W_{\mathcal{C}_0}^{\mbox{\tiny
H}}(x,y)$, and for $s \geq 0$ let $W_{\mathcal{C}_s}^{\mbox{\tiny
H}}(x,y)$ be the weight enumerator of its $s$-th order coordinate
extension $\mathcal{C}_s$. Then
\begin{equation}\label{eq:wh_Ws}
    W_{\mathcal{C}_s}^{\mbox{\tiny H}}(x,y) = W_{\mathcal{C}_0}^{\mbox{\tiny H}}(x,y)\cdot\left[x+(q^m-1)y\right]^s.
\end{equation}
\end{lemma}

Combining Corollary~\ref{cor:wh_dual_vector},
Lemma~\ref{lemma:wh_MDS}, and Lemma~\ref{lemma:wh_Ws}, the Hamming
weight distribution of $\mathcal{L}$ can eventually be determined.

\begin{proposition}\label{prop:wh_W_L}
For ${\bf v} \in \mathrm{GF}(q^m)^n$ with $\wh({\bf v}) = r$, the
Hamming weight enumerator of $\mathcal{L} = \vspan{{\bf v}}^{\perp}$
depends on only $\wh({\bf v})$, and is given by
\begin{eqnarray}
    \nonumber
    W_\mathcal{L}^{\mbox{\tiny H}}(x,y) &=& q^{-m} \Big\{ \left[x+(q^m-1)y\right]^{n} + (q^m-1) \cdots\\
    & \cdots & (x-y)^{r} \left[x+(q^m-1)y\right]^{n-r} \Big\}.
\end{eqnarray}
\end{proposition}

\begin{lemma}\label{lemma:f_H_hat}
Suppose ${\bf v} \in \mathrm{GF}(q^m)^n$ has Hamming weight $r$.
Then the Hadamard transform of the Hamming weight function is given
by
\begin{equation}\label{eq:f_wh}
    \hat{f}_{\mbox{\tiny{H}}}({\bf v} ) = (x-y)^{r}[x+(q^m-1)y]^{n-r}.
\end{equation}
\end{lemma}

Using Lemma~\ref{lemma:f_H_hat}, we finally establish the
MacWilliams identity for the Hamming metric.

\begin{theorem}\label{th:hamming_macw}
For any linear code $\mathcal{C}$, we have
\begin{equation}
    W_{\mathcal{C}^{\perp}}^{\mbox{\tiny H}}(x,y) = \frac{1}{|\mathcal{C}|}
    W_\mathcal{C}^{\mbox{\tiny H}}\left(x+(q^m-1)y,x-y\right).
\end{equation}
\end{theorem}
We remark that the MacWilliams identities for the Hamming and the
rank metrics given in Theorems~\ref{th:hamming_macw} and
\ref{th:MacWilliams} respectively have exactly the same form except
for the $q$-transform in Eq.~(\ref{eq:macwilliams}). Note that
Theorem~\ref{th:hamming_macw} is precisely the MacWilliams identity
for the Hamming metric given by Theorem~13 in
\cite[Chap.~5]{macwilliams_77}, although our proof is different from
that in \cite[Chap.~5]{macwilliams_77}. Finally, we remark that
Theorem~13 in \cite[Chap.~5]{macwilliams_77} is a special case of
the MacWilliams Theorem for complete weight enumerators (see
Theorem~10 in \cite[Chap.~5]{macwilliams_77}). For the rank metric,
it is not clear how we can adapt the concept of complete weight
enumerator to give a proof of the MacWilliams identity.

\bibliographystyle{IEEETran}
\bibliography{gpt}

\end{document}